# Geometric phase in the $G_3^+$ quantum state evolution

Alexander SOIGUINE[1]

[1] SOiGUINE Supercomputing, Aliso Viejo, CA 92656, USA

**Email address:**
alex@soiguine.com

**Key words: Qubits, Geometric algebra, Clifford translation, Geometric phase, Topological quantum computing**

**Abstract:** When quantum mechanical qubits as elements of two dimensional complex Hilbert space are generalized to elements of even subalgebra of geometric algebra over three dimensional Euclidian space, geometrically formal complex plane becomes explicitly defined as an arbitrary, variable plane in 3D [1]. The result is that the quantum state definition and evolution receive more detailed description, including clear calculations of geometric phase, with important consequences for topological quantum computing.

## 1. Introduction

Qubits, unit value elements of the Hilbert space $C^2$ of two dimensional complex vectors:

$$|\psi\rangle = \begin{pmatrix} z_1 \\ z_2 \end{pmatrix} = z_1 \begin{pmatrix} 1 \\ 0 \end{pmatrix} + z_2 \begin{pmatrix} 0 \\ 1 \end{pmatrix}, \quad z_1^2 + z_2^2 = z_1 \tilde{z}_1 + z_2 \tilde{z}_2 = 1, \quad z_k = z_k^1 + i z_k^2, \quad k = 1,2$$

can be generalized to unit value elements $\alpha + I_S \beta$ of even subalgebra $G_3^+$ of geometric algebra $G_3$ over Euclidian space $E_3$, g-qubits [1]:

$$\alpha + I_S \beta = \alpha + \beta(b_1 B_1 + b_2 B_2 + b_3 B_3) = \alpha + \beta_1 B_1 + \beta_2 B_2 + \beta_3 B_3, \beta_i = \beta b_i, \ \beta_i = \beta b_i, \ \alpha^2 + \beta^2 = 1,$$

$$b_1^2 + b_2^2 + b_3^2 = 1, \quad \alpha, \beta, b_i \text{ scalars}, \ B_i \text{ - unit value bivectors satisfying}$$

$$B_1 B_2 = -B_3, \quad B_1 B_3 = B_2, \quad B_2 B_3 = -B_1, \quad B_i^2 = -1 \quad (1.1)$$

Multiplication rules (1.1) assume the right screw space orientation that can be seen through the order of vectors, dual to the bivectors, used to create the space oriented unit volume $I_3$ (see Fig.1.1).

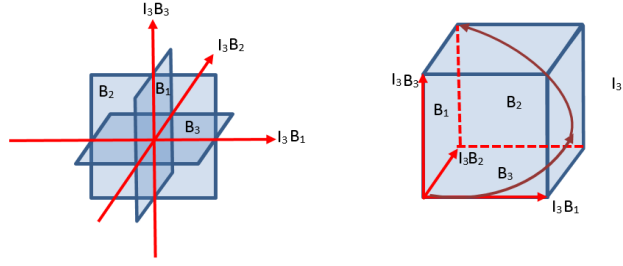

Fig.1.1. Right screw unit value oriented volume

There always are two options to create oriented unit volume, depending on the order of vectors in the product. They correspond to the two types of the three dimensional space handedness – left and right screw handedness. One can also think about $I_3$ as a right (left) single thread screw helix of the height one (see the above picture). In this way $-I_3$ is left (right) screw helix.

Mappings between g-qubits and qubits are not one-to-one and are defined by:

$$\left.\begin{array}{l} so(\alpha,\beta,S) = \alpha + \beta_1 B_1 + \beta_2 B_2 + \beta_3 B_3 = \\ \alpha + \beta_1 B_1 + (\beta_3 + \beta_2 B_1) B_3 = \\ \alpha + \beta_2 B_2 + (\beta_1 + \beta_3 B_2) B_1 = \\ \alpha + \beta_3 B_3 + (\beta_2 + \beta_1 B_3) B_2 \end{array}\right\} \Rightarrow \begin{cases} \begin{pmatrix} \alpha+i\beta_1 \\ \beta_3+i\beta_2 \end{pmatrix}, i \leftarrow B_1 \\ \begin{pmatrix} \alpha+i\beta_2 \\ \beta_1+i\beta_3 \end{pmatrix}, i \leftarrow B_2 \\ \begin{pmatrix} \alpha+i\beta_3 \\ \beta_2+i\beta_1 \end{pmatrix}, i \leftarrow B_3 \end{cases} \quad (1.2)$$

that actually defines principal fiber bundle $so(\alpha,\beta,S) \xrightarrow{B_i} (z_1^i, z_2^i)$ where $\{so(\alpha,\beta,S) = g \in G_3^+ : |g|=1\}$ is total space and $\{z=(z_1,z_2) \in C^2; |z|^2 =1\}$ is base space. I will denote them as $G_3^+|_{S^3}$ and $C^2|_{S^3}$ respectively. The projection $\pi: G_3^+|_{S^3} \to C^2|_{S^3}$ depends on which particular $B_i$ is taken from an arbitrarily selected triple $\{B_1, B_2, B_3\}$ in 3D satisfying (1.1). Bivector $B_i$ defines complex plane for the complex vectors of $C^2$, so we should write $\pi: G_3^+|_{S^3} \xrightarrow{B_i} C^2|_{S^3}$. For any $z = \begin{pmatrix} x_1+iy_1 \\ x_2+iy_2 \end{pmatrix} \in C^2|_{S^3}$ the fiber in $G_3^+|_{S^3}$ consists of all elements $F_z = x_1 + y_2 B_1 + x_2 B_2 + y_1 B_3$, if $B_3$ is optionally chosen as complex plane. That particularly means that standard fiber is equivalent to the group of rotations of the triple $\{y_2 B_1, x_2 B_2, y_1 B_3\}$ as a whole. All such rotations in $G_3^+$ are also identified by elements of $G_3^+|_{S^3}$ since for any bivector $B$ the result of its rotation is [1] (see, for example

---

[1] It is convenient to write elements $so(\alpha,\beta,S)$ as exponents: $so(\alpha,\beta,S) = e^{Is\varphi}$.

[2], [3]): $so(\gamma, \delta, S)\tilde{}\ Bso(\gamma, \delta, S)$, where $so(\gamma, \delta, S)\tilde{} = \gamma - \delta_1 B_1 - \delta_2 B_2 - \delta_3 B_3$. So, standard fiber is identified as $G_3^+|_{S^3}$ and the composition of rotations is:

$$e^{-I_{S_2}\psi}\left(e^{-I_{S_1}\varphi} B e^{I_{S_1}\varphi}\right) e^{I_{S_2}\psi} = \left(e^{I_{S_1}\varphi} e^{I_{S_2}\psi}\right)\tilde{}\ B e^{I_{S_1}\varphi} e^{I_{S_2}\psi}$$

Multiplications of $so(\alpha, \beta_1, \beta_2, \beta_3) = \alpha + \beta_1 B_1 + \beta_2 B_2 + \beta_3 B_3$ by basis bivectors $B_i$ give basis bivectors of tangent spaces to original bivectors [4]:

$T_1 = so(\alpha, \beta_1, \beta_2, \beta_3) B_1 = (\alpha + \beta_1 B_1 + \beta_2 B_2 + \beta_3 B_3) B_1 = -\beta_1 + \alpha B_1 - \beta_3 B_2 + \beta_2 B_3$

$T_2 = so(\alpha, \beta_1, \beta_2, \beta_3) B_2 = (\alpha + \beta_1 B_1 + \beta_2 B_2 + \beta_3 B_3) B_2 = -\beta_2 + \beta_3 B_1 + \alpha B_2 - \beta_1 B_3$

$T_3 = so(\alpha, \beta_1, \beta_2, \beta_3) B_3 = (\alpha + \beta_1 B_1 + \beta_2 B_2 + \beta_3 B_3) B_3 = -\beta_3 - \beta_2 B_1 + \beta_1 B_2 + \alpha B_3$

These $G_3^+|_{S^3}$ elements are orthogonal to $\alpha + \beta_1 B_1 + \beta_2 B_2 + \beta_3 B_3$ and to each other, and are the tangent space basis elements at points $\alpha + \beta_1 B_1 + \beta_2 B_2 + \beta_3 B_3$.

Projections of $T_i$ onto $C^2|_{S^3}$ are:

$$\pi(T_1) = \begin{pmatrix} -\beta_1 + i\beta_2 \\ -\beta_3 + i\alpha \end{pmatrix}, \quad \pi(T_2) = \begin{pmatrix} -\beta_2 - i\beta_1 \\ \alpha + i\beta_3 \end{pmatrix}, \quad \pi(T_3) = \begin{pmatrix} -\beta_3 + i\alpha \\ \beta_1 - i\beta_2 \end{pmatrix}$$

These elements of $C^2|_{S^3}$ are mutually orthogonal, in the sense of Euclidean scalar product in $C^2$:

$\langle z, w \rangle = \text{Re}(\tilde{z}_1 w_1 + \tilde{z}_2 w_2)$, and orthogonal to the projection $\pi(so(\alpha, \beta_1, \beta_2, \beta_3)) = \begin{pmatrix} \alpha + i\beta_3 \\ \beta_2 + i\beta_1 \end{pmatrix}$ of the original state in $G_3^+|_{S^3}$. They are the tangent space basis elements in $C^2|_{S^3}$ at points $\begin{pmatrix} z_1 \\ z_2 \end{pmatrix} = \begin{pmatrix} \alpha + i\beta_3 \\ \beta_2 + i\beta_1 \end{pmatrix}$

.

2. Clifford translations

Let's take Clifford translation in $C^2|_{S^3}$, $Cl_\psi(z) = e^{i\psi} z$, and lift it to $G_3^+|_{S^3}$ using:

$$F_{Cl_\psi(z)} = r_1 \cos(\psi + \psi_1) + r_2 \sin(\psi + \psi_2) B_1 + r_2 \cos(\psi + \psi_2) B_2 + r_1 \sin(\psi + \psi_1) B_3 =$$
$$r_1 \cos(\psi + \psi_1) + r_1 \sin(\psi + \psi_1) B_3 + [r_2 \cos(\psi + \psi_2) + r_2 \sin(\psi + \psi_2) B_3] B_2 =$$
$$r_1 e^{B_3 \psi} e^{B_3 \psi_1} + r_2 e^{B_3 \psi} e^{B_3 \psi_2} B_2 = e^{B_3 \psi} F_z$$

Translational velocity is

$$\frac{\partial}{\partial \psi} F_{Cl_\psi(z)} = \frac{\partial}{\partial \psi}\left(e^{B_3\psi} F_z\right) = B_3 F_{Cl_\psi(z)} \quad (2.1)$$

and is orthogonal to $F_{Cl_\psi(z)}$:

$$\left\langle F_{Cl_\psi(z)}, B_3 F_{Cl_\psi(z)} \right\rangle_0 = \left(F_{Cl_\psi(z)} \tilde{F}_{Cl_\psi(z)} \tilde{B}_3\right)_0 = (B_3)_0 = 0 \text{ (index 0 means scalar part of element)}$$

Two other components of the tangent space, orthogonal to $F_{Cl_\psi(z)}$ and $B_3 F_{Cl_\psi(z)}$ at any point of the orbit, are $B_1 F_{Cl_\psi(z)}$ and $B_2 F_{Cl_\psi(z)}$[2]. Their velocities, while moving along Clifford orbit are:

$$\frac{\partial}{\partial \psi}\left(B_1 F_{Cl_\psi(z)}\right) = \frac{\partial}{\partial \psi}\left(B_1 e^{B_3\psi} F_z\right) = B_1 B_3 F_{Cl_\psi(z)} = B_2 F_{Cl_\psi(z)} \quad (2.2)$$

(derivative of $B_1 F_{Cl_\psi(z)}$ is orthogonal to $B_1 F_{Cl_\psi(z)}$ and looking in the direction $B_2 F_{Cl_\psi(z)}$)

$$\frac{\partial}{\partial \psi}\left(B_2 F_{Cl_\psi(z)}\right) = \frac{\partial}{\partial \psi}\left(B_2 e^{B_3\psi} F_z\right) = B_2 B_3 F_{Cl_\psi(z)} = -B_1 F_{Cl_\psi(z)} \quad (2.3)$$

(derivative of $B_2 F_{Cl_\psi(z)}$ is orthogonal to $B_2 F_{Cl_\psi(z)}$ and looking in the direction $-B_1 F_{Cl_\psi(z)}$)

These two equations explicitly show that the two tangents, orthogonal to Clifford translation velocity, rotate in moving plane $\{B_1 F_{Cl_\psi(z)}, B_2 F_{Cl_\psi(z)}\}$ with the same, by value, rotational velocity as translation velocity is (see Fig 2.1).

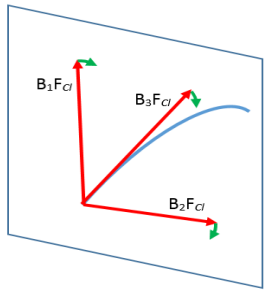

Fig.2.1. Tangents rotate in their plane with the same, by value, speed as translational velocity

---

[2] Clearly, the three $B_i F_{Cl_\psi(z)}$ are identical to earlier considered tangents $T_i$

If a fiber, g-qubit, makes full circle in Clifford translation: $F_z \rightarrow F_{Cl_\psi(z)} = e^{B_3\psi}F_z$, $0 \leq \psi \leq 2\pi$, both $B_1 F_{Cl_\psi(z)}$ and $B_2 F_{Cl_\psi(z)}$ also make full rotation in their common plane by $2\pi$. This is special case of the g-qubit geometric phase, incrementing in the $S^3$ sphere big circle closed curve quantum state path.

The demonstrated rotation of tangents in the plane orthogonal to the orbit of g-qubit Clifford translation – that is what actually is not intuitively obvious and is more important than all widely accepted "mysteries" of quantum mechanics. This rotation phenomena has nothing to do with the size of physical system. This is topological property of the space of dimension 4, not 3, our imagination cannot easily deal with.

At the same time, we should remember that g-qubits, states, are operators acting on observables. Though observables are elements of the same space as states (see next section), action of a state on observable is $G_3^+ \xrightarrow{G_3^+} G_3^+ : \alpha + I_S\beta \xrightarrow{C} (\alpha - I_S\beta)C(\alpha + I_S\beta)$, and the result of this action changes differently compared to the state modification subjected to Clifford translation.

### 3. Measurement of observables in basis states

Let's consider the case when, for an arbitrary g-qubit $\alpha + I_S\beta$, the plane $I_S$ is taken as playing the role of complex plane. Then, due to (1.2), the $C^2|_{S^3}$ element, given by projection $so(\alpha,\beta,S) \xrightarrow{S} (z_1, z_2)$, is $\begin{pmatrix} \alpha + i\beta \\ 0 \end{pmatrix}, i = S$.

Let's recall the definitions of states, observables and measurements, appropriate for the case of the $G_3^+$ formalism of the two state systems [1].

States and observables are elements of $G_3^+$:

*Definition 3.1 (state, unit value element of $G_3^+$, defines operation acting on observable in a measurement ):*

$$so(\alpha, \beta, S) \equiv \alpha + I_S\beta = \alpha + \beta(b_1 B_1 + b_2 B_2 + b_3 B_3) = \alpha + \beta_1 B_1 + \beta_2 B_2 + \beta_3 B_3, \ \beta_i = \beta b_i,$$
$$\alpha^2 + \beta^2 = 1, \ b_1^2 + b_2^2 + b_3^2 = 1$$

*Definition 3.2 (observable, element of $G_3^+$):*

$$C = C_0 + C_1 B_1 + C_2 B_2 + C_3 B_3$$

*Definition 3.3 (measurement):* Measurement of observable $C$, measured in state $\alpha + I_S \beta$, is generalized Hopf fibration generated by the observable:

$$G_3^+ \xrightarrow{G_3^+} G_3^+ : \alpha + I_S \beta \xrightarrow{C} (\alpha - I_S \beta) C (\alpha + I_S \beta)$$

Explicit formulas can be found in [1].

Due to definition (3.1) the state $\alpha + I_S \beta$ corresponds to the "stable state" (see [5]), $|0\rangle$ in familiar terms of quantum mechanical notations $|\psi\rangle = z_1 |0\rangle + z_2 |1\rangle$, if $I_S$ is selected as complex plane. The state $\alpha + I_S \beta$ is "stable" in the sense that the measurement of any observable with the bivector part parallel to $I_S$ does not change the observable (I omit scalar part which does not change in rotations):

$$(\alpha - I_S \beta)\gamma I_S (\alpha + I_S \beta) = (\alpha \gamma I_S + \beta \gamma)(\alpha + I_S \beta) = \alpha^2 \gamma I_S + \alpha \beta \gamma - \alpha \beta \gamma + \beta^2 \gamma I_S =$$
$$(\alpha^2 + \beta^2)\gamma I_S = \gamma I_S \quad (3.1)$$

The g-qubit state corresponding to $|1\rangle$ is $\beta_2 I_{S_2} + \beta_3 I_{S_3}$, $\beta_2^2 + \beta_3^2 = 1$, where $I_{S_2}$ is any unit bivector orthogonal to $I_S$ and $I_{S_3} = I_{S_2} I_S$ (if we keep right screw space orientation with multiplication rules (1.1).) Measurement of any observable with the bivector part parallel to $I_S$ in this state gives:

$$(-\beta_2 I_{S_2} - \beta_3 I_{S_3})\gamma I_S (\beta_2 I_{S_2} + \beta_3 I_{S_3}) = (-\beta_2 \gamma I_{S_3} + \beta_3 \gamma I_{S_2})(\beta_2 I_{S_2} + \beta_3 I_{S_3}) =$$
$$-\beta_2^2 \gamma I_S - \beta_2 \beta_3 \gamma + \beta_2 \beta_3 \gamma - \beta_3^2 \gamma I_S = -(\beta_2^2 + \beta_3^2)\gamma I_S = -\gamma I_S \quad (3.2)$$

The last formula means that measurement of any observable with the bivector part parallel to $I_S$, in the state corresponding to $|1\rangle$, flips bivector part of the observable.

Formulas (3.1), (3.2) retrieve the actual sense of the two basis states.

Consider the results of measurements in states $\alpha + I_S \beta$ and $\beta_2 I_{S_2} + \beta_3 I_{S_3}$ of an arbitrary observable $C = C_0 + C_1 I_S + C_2 I_{S_2} + C_3 I_{S_3}$:

$$(\alpha - \beta I_S)C(\alpha + \beta I_S) = C_0 + C_1 I_S + [(\alpha^2 - \beta^2)C_2 - 2\alpha\beta C_3] I_{S_2} + [(\alpha^2 - \beta^2)C_3 + 2\alpha\beta C_2] I_{S_3} =$$
$$C_0 + C_1 I_S + (C_2 \cos 2\varphi - C_3 \sin 2\varphi) I_{S_2} + (C_2 \sin 2\varphi + C_3 \cos 2\varphi) I_{S_3} \quad (3.3)$$

(through parameterization $\alpha = \cos\varphi$, $\beta = \sin\varphi$)

$$(-\beta_2 I_{S_2} - \beta_3 I_{S_3})C(\beta_2 I_{S_2} + \beta_3 I_{S_3}) = C_0 - C_1 I_S + [C_2(\beta_2^2 - \beta_3^2) + 2C_3 \beta_2 \beta_3] I_{S_2} + [2C_2 \beta_2 \beta_3 - C_3(\beta_2^2 - \beta_3^2)] I_{S_3} =$$
$$C_0 - C_1 I_S + (C_2 \cos 2\vartheta + C_3 \sin 2\vartheta) I_{S_2} + (C_2 \sin 2\vartheta - C_3 \cos 2\vartheta) I_{S_3} \quad (3.4)$$

(through parameterization $\beta_2 = \cos\vartheta$, $\beta_3 = \sin\vartheta$).

Formulas (3.3), (3.4) mean the following:

Measurement of observable $C = C_0 + C_1 I_S + C_2 I_{S_2} + C_3 I_{S_3}$ in pure qubit state $\alpha + I_S \beta$ has bivector part with the $I_S$ component equal to unchanged value $C_1$. The $I_{S_2}$ and $I_{S_3}$ measurement components are equal to $I_{S_2}$ and $I_{S_3}$ components of $C$ rotated by angle $2\varphi$ defined by $\alpha = \cos\varphi$ and $\beta_1 = \sin\varphi$, where plane of rotation is $I_S$.

Measurement of observable $C = C_0 + C_1 I_S + C_2 I_{S_2} + C_3 I_{S_3}$ in pure qubit state $\beta_2 I_{S_2} + \beta_3 I_{S_3}$ has bivector part with the $I_S$ component equal to flipped value $-C_1$ (flipping in $I_S$ plane). The $I_{S_2}$ and $I_{S_3}$ measurement components are equal to $I_{S_2}$ and $I_{S_3}$ components of $C$ rotated by angle $2\vartheta$ defined by $\beta_2 = \cos\vartheta$, $\beta_3 = \sin\vartheta$, where plane of rotation is $I_S$. The absolute value of angle of rotation is the same as for $\beta_3 + I_S \beta_2$ but the rotation direction is opposite to the case of $\beta_3 + I_S \beta_2$.

The above two results are geometrically pretty clear. The two states $\beta_3 + I_S \beta_2$ and $\beta_2 I_{S_2} + \beta_3 I_{S_3} = (\beta_3 + I_S \beta_2) I_{S_3}$ only differ by additional factor $I_{S_3}$ in $(\beta_3 + I_S \beta_2) I_{S_3}$. That means that measurements of an observable $C$, if it is pure bivector, in states $\beta_3 + I_S \beta_2$ and $\beta_2 I_{S_2} + \beta_3 I_{S_3}$ are equivalent up to additional "wrapper" $I_{S_3}$:

$$((\beta_3 + I_S \beta_2) I_{S_3})\tilde{\ } C(\beta_3 + I_S \beta_2) I_{S_3} = \tilde{I}_{S_3} (\beta_3 + I_S \beta_2)\tilde{\ } C(\beta_3 + I_S \beta_2) I_{S_3}$$

That simply means that the measurement in state corresponding to $|1\rangle$ is received from the $(\beta_3 + I_S \beta_2)\tilde{\ } C(\beta_3 + I_S \beta_2)$ measurement, measurement in state corresponding to $|0\rangle$, just by mirroring the result relative to the plane $I_{S_3}$ (see Fig.3.1).

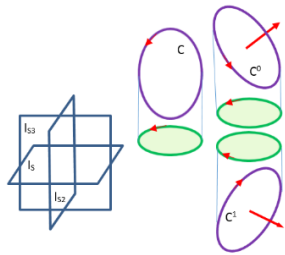

Fig.3.1. Results of measurement of C in the two stable states. C⁰, any measurement in a state corresponding to $|0\rangle$, and C¹, other measurement in a state corresponding to $|1\rangle$, can be made mirrored of each other by rotating in a plane parallel to Is. The Is components are the same, in absolute value, for all the three C, C⁰ and C¹.

## 4. Sequence of infinitesimal Clifford translations

Since any bivector in $G_3^+$ can be generally taken as playing the role of "complex" plane, let's take some arbitrary unit bivector $I_{S_3}$ [3] and make infinitesimal Clifford translation of an arbitrary g-qubit, state:

$$so(\alpha, \beta, S) \to e^{I_{S_3} d\varphi} so(\alpha, \beta, S) \equiv Cl(I_{S_3}, d\varphi)(so(\alpha, \beta, S))$$

Instant translational velocity tangent of it is: $I_{S_3} Cl(I_{S_3}, d\varphi)(so(\alpha, \beta, S))$. We also need two bivectors (planes) to create rotational tangent components. The first one, $I_{S_1}$, is any unit bivector orthogonal to $I_{S_3}$, so defined up to arbitrary angle of rotation around normal to $I_{S_3}$. The bivector for the second tangent can be taken, in the case of the right screw space orientation, as $I_{S_2} = I_{S_1} I_{S_3}$.

The translational velocity tangent will rotate by the value $d\varphi$ in the direction opposite to $Cl(I_{S_3}, d\varphi)(so(\alpha, \beta, S))$ because from (2.1):

$$\frac{\partial}{\partial \psi} I_{S_3} F_{Cl_\psi(z)} = I_{S_3} I_{S_3} F_{Cl_\psi(z)} = -F_{Cl_\psi(z)}$$

The rotational velocity tangents will rotate by the same $d\varphi$ value in their plane orthogonal to translational velocity tangent as normal (see Fig. 2.1, replacing $B_i$ to $I_{S_i}$). All that means that while moving in a sequence of infinitesimal Clifford translations the two rotational tangents rotate in each infinitesimal step by the same angle in their plane as translational velocity tangent rotates in plane $I_{S_3}$ moving along orbit lying on $S^3$.

We saw above that it does not matter do we look at the angle accumulated by translational velocity in the varying plane spanned by this velocity together with instant state g-qubit, or accumulated angle of rotation of any of the two tangents of rotational speed - they are equal!

Let we move along some arbitrary path on $S^3$. The sphere is center symmetrical surface, so at any instant value of state on $S^3$ infinitesimal incrementing of translational velocity angle does not depend in what direction the displacement happens. If the path is approximated with infinitesimal pieces of geodesics then the accumulated angle between translational velocity and instant geodesic is obviously equal to the total length of the path (see Fig.4.2).

---

[3] Index "3" here is taken to stress that this bivector will play the same role as the previously used $B_3$ in Sec. 3. Hopefully, reader remembers that $S_3$ is a plane in 3D bearing index "3", and $S^3$ is unit 3-dimensional sphere in 4-dimensional space.

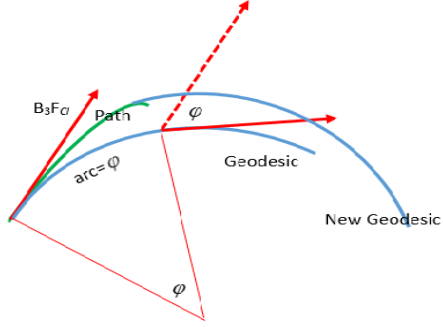

Fig.4.2. Accumulating of angle while moving along path. Also equals by value to rotation angle of two vectors in plane remaining orthogonal to translational velocity

This is purely geometrical phase, separated from g-qubit exponent phase modifications caused by external factors defining the g-qubit (state) path on the $S^3$. Infinite composition of infinitesimal Clifford translations along any $S^3$ path $L$ with varying $I_{S_3}$ gives the final state g-qubit.

Take a sequence of infinitesimal Clifford transformations:

$$e^{I_{S_3(l_N)}\Delta l_N}...e^{I_{S_3(l_2)}\Delta l_2}e^{I_{S_3(l_1)}\Delta l_1}so(\alpha,\beta,S)$$

By taking the logarithm, approaching $N \to \infty$ and getting back to exponent we receive the final state:

$$\int_L e^{I_{S_3(l)}dl}so(\alpha,\beta,S) \quad (4.1)$$

## 5. Conclusions

Evolution of a quantum state described in terms of $G_3^+$ gives more detailed information about two state system compared to the $C^2$ Hilbert space model. It confirms the idea that distinctions between "quantum" and "classical" states become less deep if a more appropriate mathematical formalism is used. The paradigm spreads from trivial phenomena like tossed coin experiment [6] to recent results on entanglement and Bell theorem [7] where the former was demonstrated as not exclusively quantum property.